# Role of pinning mechanism in co-precipitation derived cobalt rich, cobalt ferrite nanoparticles

Sai Srikanth Arvapalli[1], Bhusankar Talluri[1,2], Kousika Anbalagan[1], and Tiju Thomas[1*]

[1]Department of Metallurgical and Material Engineering, and [2]Department of Chemistry, Indian Institute of Technology Madras, Chennai-600036, India.

*Correspondence:*

* E-mail: tijuthomas@iitm.ac.in, tt332@cornell.edu; Fax: +91-44- 2257-4752; Tel: +91- 44- 2257-5781 (Lab)

## Abstract

Cobalt ferrite nanoparticles have been synthesized using a co-precipitation approach. Co:Fe precursor ratio is varied over a certain window (1.05 to 1.2). It is observed that the 1.15:2 precursor ratio gives better phase control but poor coercivity. On the other hand, 1.05:2 precursor ratio results in substantially better coercivity values (274kA/m; almost 300% the value reported for co-precipitated cobalt ferrite by Praveena et al.), but moderate BH product maximum (2.25 kJ/m$^3$; ~ comparable to many reports on cobalt ferrite nanoparticles so far). The nanoparticles with best coercivity are annealed at 873K for varying durations (2, 4, 6, 12 hrs). It is observed that the coercivity drops drastically (almost by 80%) after annealing for 2 hours. However thereafter coercivity and saturation magnetization improves marginally with increasing duration of annealing. These studies, along with thermogravimetric analysis, and infrared spectroscopic results indicate that a hydroxide nanophase based flux pinning mechanism at the grain boundary plays an important role in explaining the observed magnetic property trends. It is believed that this result will be generically helpful in developing soft chemically derived ferrites with higher coercivity and moderate (BH)$_{max}$. However to develop plausible applications using the reported ferrites that use nanophase flux pinning, soft materials and device processing methods will need to be explored.

1. **Introduction**

Cobalt ferrite ($CoFe_2O_4$) is a ferrimagnetic material with tunable coercivity (1.67x103 - 4.65x105 A/m) [1], moderate magnetization (around 80$Am^2$/kg) and good thermal and chemical stability [2]. It is being used in several technological applications such as computer memory core elements and switches, magnetic recording media [3], core of coils in microwave frequency devices, contrast agents for magnetic resonance imaging (MRI) [4], probes for magnetic force microscopy, hyperthermia based tumour treatment [5], drug delivery [6], and magnetically recoverable catalyst [7-9]. Crystallographically, the ferrite spinel structure ($AB_2O_4$) is a closed-packed oxygen lattice, in which tetrahedral (A sites) and octahedral (B sites) interstices are occupied by the cations. In the inverse spinel structure, all the $Co^{2+}$ ions occupy the octahedral sites of the lattice. Half of the $Fe^{3+}$ ions occupy the octahedral sites and the rest of the $Fe^{3+}$ ions are present in tetrahedral sites [10].

Large-scale industrial applications of nanoscale ferrites have motivated the development of widely used chemical methods, including solution combustion [11], co-precipitation [12], sol-gel [13] and precipitation [2] for the synthesis of stoichiometric and chemically pure spinel ferrite nanoparticles. Magnetization can be varied by introducing off-stoichiometry (like oxygen vacancies and excess metal ions) in the lattice structure. Controlling the stoichiometry in turn can be achieved by varying the synthesis conditions [12]. Co-precipitation is a simple and green synthesis method for producing nanoparticles. The advantage of this method is that it offers better size control; the polydispersity is largely controlled through the handle on the relative rates of nucleation and growth during the synthesis process [14]. Size control is achieved by the addition of a precipitating agent (eg. NaOH) to the precursor solution; vigorous stirring is usually employed in this process [2].

It may be noted that the coercivity of a magnetic material depends on the particle dimensions (in addition to composition). Hence particle size control and appropriate grain boundary processing is critical for obtaining optimum coercivity. In this work, we explore (i) precursor ratio, and (ii) annealing temperature and time in obtaining optimal coercivity. Results indicate that nanophases undetectable via x-ray diffraction seem to be responsible for the observed coercivity in the as prepared pristine samples (with Co:Fe=1.05:2). Any heat treatment affects this coercivity, indicating the interfacial origins of pinning and hence coercivity enhancement in this sample.

2. **Experimental section**

To study the effect of off-stoichiometry on the magnetic properties, solutions with 0.21, 0.22, 0.23, 0.24 M cobalt nitrate (50ml) and 0.4M (200ml) ferric nitrate are prepared separately using de-ionised water. The suitably chosen cobalt nitrate solution is mixed with the iron containing one (50ml:50ml). A particular pH range (approx. 11.7) is maintained by the addition of 2M sodium hydroxide. Higher pH is chosen to ensure fast growth kinetics, which in turn results in larger crystallite sizes [12]. This precipitate is collected, washed several times to remove soluble impurities and pelletized. Products made using different precursor concentration are labelled accordingly (ref: Table 1).

Sample with best coercivity value (274.7KA/m) is annealed at 873K for varied durations (2, 4, 6, and 12 hours). The sintering temperature is chosen based on the observations made by Praveena et.al. [12], in order to obtain cobalt ferrite nanoparticles with moderate Fano factor ($\sigma^2/d$, where $\sigma$-variance and d= mean of Average crystallite size). Fano factor is a statistical measure of size dispersion in nanoparticles. As the calcination temperature increases, the crystallite size increases while the Fano factor decreases [12]. Therefore to obtain ferrite with good magnetization values as well as relative monodispersity, a moderate annealing temperature (873K) is chosen.

To characterize pristine and sintered samples, PANalyticalX'pert powder X-ray diffractometer (XRD) is used with Cu $K_\alpha$ (1.54A) radiation. The morphology and composition are examined using Energy Dispersive X-ray Spectroscopy (EDS) and Scanning Electron Microscope (Inspect F SEM). Room temperature measurements are made using Vibrating Sample Magnetometer (Lakeshore VSM 7410). Weight loss measurements are done using Thermogravimetric analyser (SDT Q600 V20.9 Build 20). FTIR analysis is done using Perkin Elmer Spectrum1 FT-IR instrument. Reitveld analysis is carried out using MAUD (Materials Analysis Using Diffraction) software..

**Table 1:** Sample labels used in this work

| Precursor ratio (Fe : Co) | Sample name (Product) |
|---|---|
| 2 : 1.05 | $C_{1.05}$ |
| 2 : 1.1 | $C_{1.1}$ |
| 2 : 1.15 | $C_{1.15}$ |
| 2 : 1.2 | $C_{1.2}$ |

## 3. Results and discussion

The X-ray diffractograms (ref: Fig. 1) of the samples $Co_{1.05}$, $Co_{1.1}$, $Co_{1.15}$ and $Co_{1.2}$ (ref: Table I), show the peaks for $CoFe_2O_4$ (reference code: 98-077-9266; ICSD code: 98553). The characteristic peak of ferrites (311) is seen in every sample processed [15]. This indicates the formation of cobalt ferrite nanoparticles and apparent absence of any secondary phases (discernible from XRD). The crystallite size (approximately 24 nm) of as prepared pristine samples is calculated using Scherrer formula at [311] peak. Rietveld refinement based analysis also suggests the absence of secondary phases as the refined calculated parameters as closely matching with ASTM data. For example, the refined calculated lattice parameter for co-precipitated $C_{1.05}$ is 0.84nm, which is almost equal to reported ASTM data (0.84nm) [16].

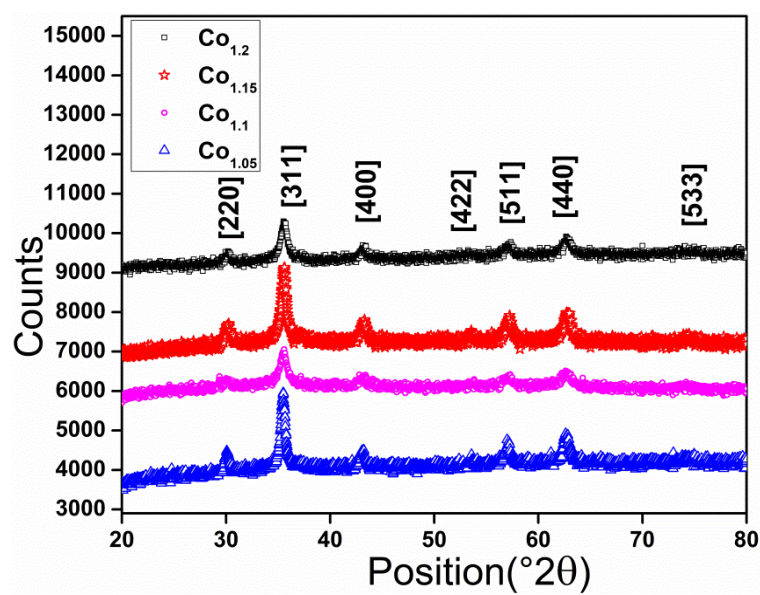

**Fig.1:** XRD pattern of the samples $Co_{1.05}$, $Co_{1.1}$, $Co_{1.15}$, and $Co_{1.2}$ confirming the formation of cobalt ferrite nanoparticles (reference code: 98-077-9266; ICSD code: 98553). [311] is observed throughout indicating cobalt ferrite phase in all samples.

**Table 2** indicates that $C_{1.05}$ shows coercivity values as high as 274.7 kA/m (ref: Fig 2); this is almost thrice the value reported by Praveena et.al. (87.9 kA/m) and is reproducibly obtained. High coercivity is observed in cobalt ferrite nanoparticle prepared by water-in-oil emulsions when precursor ratio is changed [17]. This high coercivity value is likely related to an unusual atom ratio in the sample (1:3.5 = Co:Fe; obtained from EDX/SEM). In fact this suggests the formation of a small volume fraction nanophase, which is likely segregating at the grain boundaries. Furthermore it may be expected that this large coercivity can be a result of flux pinning because of the presence of these pinning secondary nanophase (which is undetectable by SEM and XRD). This is a reasonable expectation considering prior reports [18].

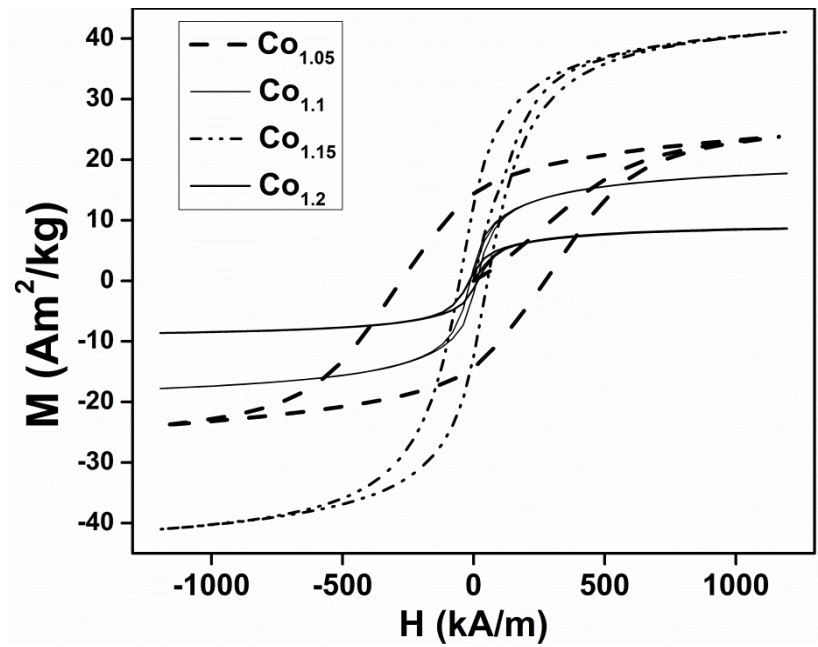

**Fig.2:** M-H hysteresis curves of $CoFe_2O_4$ with different precursor ratio ($C_{1.05}$, $C_{1.1}$, $C_{1.15}$, and $C_{1.2}$). $C_{1.05}$ shows the best $H_c$ (274.7kA/m) and $M_r$ (14.3A.m$^2$/kg) values. The coercivity of this sample is almost thrice the value previously reported by Praveena et.al. The $(BH)_{max}$ is almost 12MGOe suggesting the hard magnet properties of the cobalt ferrite nanoparticles ($C_{1.05}$).

**Table 2:** Composition of precursor, $H_c$ (coercive field), $M_s$ (saturation magnetization), $M_r$ (remnant magnetization), and Fe/Co (ratio of cations in the products) of co-precipitated samples. A substantially high $H_c$ is observed in $C_{1.05}$, indicating the fundamental difference in the sample compared to the rest.

| Composition | $H_c$ (kA/m) | $M_s$ (A.m²/kg) | $M_r$ (A.m²/kg) | Fe/Co (determined via energy dispersive X-rays) |
|---|---|---|---|---|
| $C_{1.05}$ | 274.75 | 23.8 | 14.3 | 3.5 |
| $C_{1.1}$ | 12.3 | 17.7 | 2.1 | 2.8 |
| $C_{1.15}$ | 53.1 | 41.1 | 12.5 | 2 |
| $C_{1.2}$ | 15.8 | 8.6 | 1.3 | 1.85 |

In order to confirm this "pinning by nanophase" hypothesis, annealing for several durations has been carried out. The values of coercivity dropped drastically with annealing (ref: Fig.3) as shown in **Table 3**. This drop can be due to the loss of pinning sites (plausibly due to hydroxide phases, considering the starting precursors). Interestingly this property is not seen in other co-precipitated samples ($C_{1.1}$, $C_{1.15}$, and $C_{1.2}$), indicating relevance of starting precursors (and hence the role of nucleation and kinetics of co-precipitation on obtaining high $H_c$). $C_{1.1}$, $C_{1.15}$, and $C_{1.2}$ are primarily cobalt ferrite ultrafine nanoparticles (average particle size: 17nm, 20nm, and 17nm respectively; ref: Fig. 4) which are superparamagnetic in nature (ref: Fig. 2). Hence the overall coercivity values of these samples are not high. Size distribution analysis of samples show log-normal distribution based on histogram obtained from SEM images. This is consistent with what Praveena et al. observed [12]. The average particle size of the samples $C_{1.1}$, $C_{1.15}$, and $C_{1.2}$ (around 18± 2nm) is close, but different with $C_{1.05}$ (around 25nm) suggesting the difference in growth kinetics. Also particles in samples ($C_{1.1}$, $C_{1.15}$, and $C_{1.2}$) seem to be clustered or segregated, very likely because of small particle sizes in these samples [19].

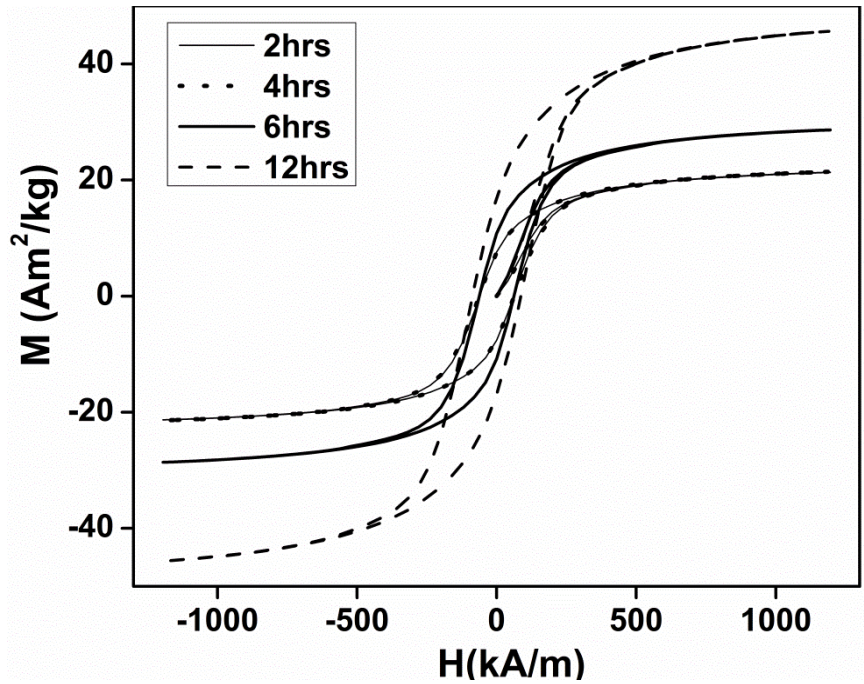

**Fig.3:** M vs H of $C_{1.05}$ sintered for 2, 4, 6, and 12 hrs at 873K. Coercivity values maintain an increasing trend with annealing duration (indicating improved magnetocrystalline anisotropy, with grain coarsening, as expected). Pinning nanophases likely reduce in volume fraction or disappear during annealing, resulting in initial coercivity loss, post annealing in $C_{1.05}$.

**Table 3:** Change in $H_c$ (coercive field), $M_s$ (saturation magnetization), and $M_r$ (remnant magnetization) with annealing durations. (*Sample studied*: $C_{1.05}$)

| Sintering Duration (Hrs) | $H_c$ (kA/m) | $M_s$ (A.m$^2$/kg) | $M_r$ (A.m$^2$/kg) |
|---|---|---|---|
| 0 (pristine sample) | 274.7 | 23.8 | 14.3 |
| 2 | 60.6 | 21.3 | 7.6 |
| 4 | 62.3 | 21.4 | 7.5 |
| 6 | 61.8 | 28.6 | 10.8 |
| 12 | 86.36 | 45.6 | 16.9 |

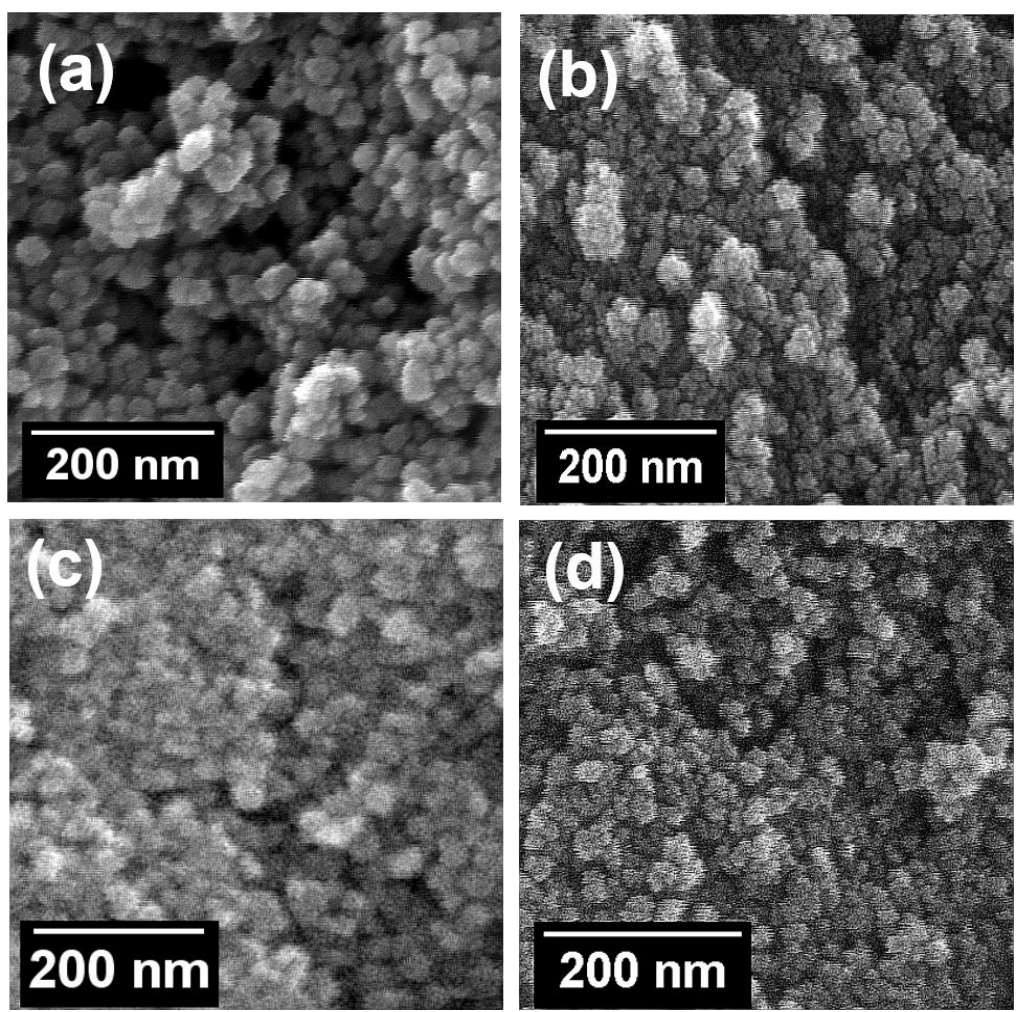

**Fig. 4**: SEM images of (**a**) $C_{1.05}$ sample. Average particle size is around 25nm which suggests that the sample is still in single domain stage. Hence upon annealing the coercivity increases with annealing duration as grain size increases. (**b, c, d**) $C_{1.1}$, $C_{1.15}$, and $C_{1.2}$ respectively. Average particle sizes of these samples are approximately 17, 20, and 17nm respectively (measured using ImageJ).

To test this hypothesis, FTIR analysis is performed (ref: Table. 4). The spectra indicate the presence of hydroxide phases in the $C_{1.05}$ (ref: Fig. 5). This is not surprising because hydroxide phases have been reported in other co-precipitated ferrites as well. These hydroxide phases are remnants of unreacted phases, which are observed and often times reported in co-precipitation of cobalt ferrite [20]. The peak observed around 1600 cm$^{-1}$ is likely to be due to stretching of $OH^{-1}$ or molecular $H_2O$, similar to the modes observed in Zn-Ni ferrite nanoparticles synthesized by co-precipitation [21]. Presence of residual nitrates (i.e. $NO_3^{-1}$ moieties) is confirmed by the 1400cm$^{-1}$ IR peak. The FTIR spectra obtained is quite broad for all the samples, as is commonly observed in inverse spinel ferrites [12]. This broadening increases with reduction of particle size, as the cation disorder is known in increase in nanoferrites [22]. This reduction in particle size [23] is responsible for increment in inverse nature of spinel ferrites (such as increment in coercivity with particle size) [24]. The main broad peaks seen around 600 cm$^{-1}$ and 3400 cm$^{-1}$ corresponds to stretching vibrations associated with metal-oxygen bonds, O-H, and N-H respectively. The peaks at about 2925, 1050 cm$^{-1}$ are due to C–H stretching and O–H bending vibrations, respectively. These indicate the presence of residual molecules adsorbed to the particle surface [25]. From Fig. 5, it can be realized that the disappearance of hydroxides due to annealing is likely responsible for the loss of coercivity. This explanation is substantiated by the broad peaks at 1600 cm$^{-1}$, 3400 cm$^{-1}$ and absence of intensity at 1050 cm$^{-1}$ in annealed sample. Hence it can be concluded that the presence of nanohydroxide phases in the particles, most likely at the interfaces, are responsible for the huge coercivity via pinning mechanism.

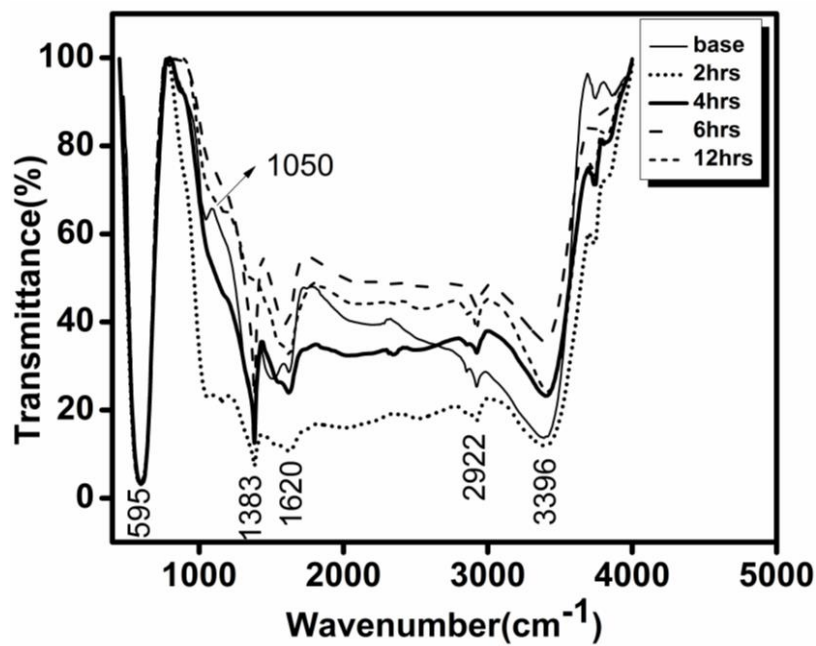

**Fig. 5:** FTIR spectrum of cobalt ferrite nanoparticles sintered for 2, 4, 6, and 12 hrs durations at 873K. The peaks at 3400 cm$^{-1}$ and 1600 cm$^{-1}$ indicates the presence of hydroxides. The intensity of these peaks reduces with annealing duration explaining the loss in high coercivity with annealing.

**Table 4:** List of peak IR positions in $C_{1.05}$

| Peak position (cm$^{-1}$) | Assignments |
|---|---|
| 600 | metal-oxygen bond vibration |
| 1050 | O-H bending |
| 1400 | $NO_3^{-1}$ |
| 1600 | $OH^{-1}$ or molecular $H_2O$ |
| 2922 | C-H stretching |
| 3395 | O-H, N-H stretching |

It may be noted that from the FTIR alone we may deduce that there are hydroxide decomposition steps occurring in these samples during annealing. Literature suggests that there would be at least two decomposition steps in the as prepared material. The first decomposition step (labelled: step I) would be for cobalt hydroxide (Co(OH)$_2$) and cobalt oxyhydroxide (CoO(OH)); these occur at around 130° and 252°C respectively [26] forming cobalt oxide (Co$_3$O$_4$) and H$_2$O [27]. The second decomposition step (labelled: step II), i.e. decomposition of cobalt oxide (Co$_3$O$_4$) formed after first decomposition step occurs at 790° and 800°C respectively [28] forming CoO and oxygen. To examine these, thermogravimetric analysis (TGA) is performed. From the TGA analysis of $C_{1.05}$ (ref: Fig. 6), it is evident that there is substantial mass loss seen over a broad range (100-200°C); indicating step-I [29]. Breadth of the TGA peak is reasonable considering compositional heterogeneities often reported in nano-ferrites [30] (and their grain boundaries) [31]. TGA shows substantial weight loss around 650°C, readily attributable to the decomposition of cobalt oxides. The small weight loss observed is due to the small volume fraction of the oxide phase at the grain boundary. Other weight losses indicated by changes in slope (ref: region-i and region-ii in Fig. 5), observed around 300 and 400°C, are likely due to decomposition of Fe-rich nitrates [29] and hydroxides [32] respectively. Since the annealing temperature is around 600°C, it can be safely assumed that the hydroxide nanophases are absent in the annealed samples. The increasing trend in coercivity of annealed samples is due to increment in particle size up to a particular value (54 ± 2nm) seen in inverse spinel ferrites as mentioned by Praveena et.al as well. As the particle size observed in the sample prepared is around 25nm (ref: Fig. 5), the coercivity increases due to increased role of magnetocrystalline anisotropy [33].

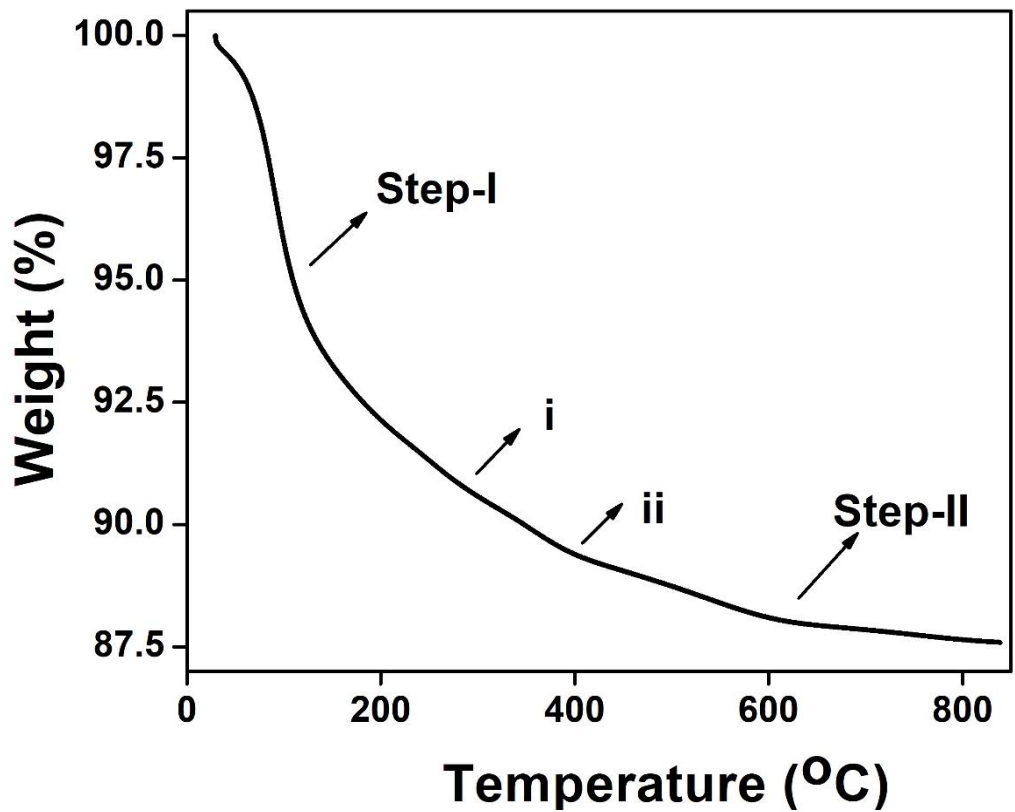

**Fig.6:** TGA analysis of $C_{1.05}$ sample indicating the weight loss step involving decomposition of cobalt hydroxides and oxides. Readily observed Steps I and II indicate weight loss due to decomposition of (I) cobalt hydroxides and $H_2O$, and (II) cobalt oxide decomposition.

Discernible changes in slope at regions – i and ii are attributed to decomposition of ferric nitrate and hydroxides.

The relevance of the results presented here may be noted in the context of the current quest for replacements for champion rare earth permanent magnets (REPMs) for novel magnetic storage devices is an important and emerging areas. Hence for novel magnetic nanomaterials, there is value in looking at the $(BH)_{max}$ parameter as well. For competing with REPMs, the $(BH)_{max}$ value must be as high as possible [34]. Here interestingly C1.05 shows high coercivity along with moderate magnetic saturation. However, only moderate $(BH)_{max}$ [35] value (almost 0.25MGOe or 2.25kJ/m3) is observed for this sample ( highest $(BH)_{max}$ observed in cobalt ferrite nanoparticles is 2.1MGOe [36-37]). This moderate $(BH)_{max}$ is because of the moderate magnetic saturation values. This result is in concurrence with previously reported high coercivity cobalt ferrite nanoparticles with $(BH)_{max}$ values ranging from 0.2-1.1MGOe [38-39]. Thus we obtain cobalt ferrite nanoparticles with both high $H_c$ as well as moderate $(BH)_{max}$ values. $(BH)_{max}$ calculations are done in SI unit system using the formula B = μ0 (H + M) [40]. Hence we suggest that $C_{1.05}$ can potentially be adapted and deployed for application in high magnetic storage applications. This result is interesting because of the green chemical nature of the co-precipitation. Hence it is believed that this result will be generically helpful in developing soft chemically/*chimie douce* (i.e low temperature chemistry) derived ferrites with higher coercivity and desirable $(BH)_{max}$. However since the coercivity drops significantly with annealing from 2 hours at 873K, to develop plausible applications using the reported ferrites that use nanophase flux pinning, soft materials and device processing methods will need to be explored.

## 4. Conclusion

High coercivity (~274 kA/m, almost 300% the previously reported value for co-precipitated cobalt ferrite nanoparticles) materials are produced via choice of a suitable precursor ratio (Co rich reaction mix; Co: Fe = 1.05:2). The product obtained is a Co rich, and apparently single phase. High $H_c$ (274kA/m) in the sample is evidently composition dependent, and concomitant with the presence of interfacial nanohydroxide phases, suggesting an interfacial pinning mechanism. This is corroborated by annealing experiments where coercivity dropped drastically (almost by 80%), while saturation magnetization increased consistently. The $(BH)_{max}$ product is moderate indicating its potential for improvement through further work and use in magnetic memory applications. Annealing past the initial coercivity drop however marginally improves both (i) coercivity and (ii) saturation magnetization indicating grain growth and associated magnetocrystalline anisotropy.

We believe that the result presented here will be generically helpful in developing soft chemically/*chimie douce* (i.e low temperature chemistry) derived ferrites with higher coercivity and sufficient $(BH)_{max}$. However since the coercivity drops significantly with annealing, to develop plausible applications using the reported ferrites that use nanophase flux pinning, soft materials and device processing methods are perhaps necessary.

**Acnowledgements**

We thank the Department of Metallurgical and Materials Engineering, Indian Institute of Technology Madras. We would like to thank the Department of Science and Technology, Government of India for support through the project nos. DST FILE NO. YSS/2015/001712 and DST 11-IFA-PH-07. We acknowledge SAIF IIT Madras, Chennai for using Vibrating Sample Magnetometer and FTIR spectrophotometer.

**References**

[1] Y. Cedeño-Mattei and O. Perales-Pérez, "Synthesis of high-coercivity cobalt ferrite nanocrystals," Microelectron. J., vol. 40, no. 4–5, pp. 673–676, May 2009.

[2] Mahboubeh Houshiar, Fatemeh Zebhi, Zahra Jafari Razi, Ali Alidoust, and Zohreh Askari, "Synthesis of cobalt ferrite (CoFe$_2$O$_4$) nanoparticles using combustion, coprecipitation, and precipitation methods: A comparison study of size, structural, and magnetic properties," J. Magn. Magn. Mater., vol. 371, pp. 43–48, Dec. 2014.

[3] Shouheng Sun et al., "Monodisperse MFe$_2$O$_4$ (M = Fe, Co, Mn) Nanoparticles," J. Am. Chem. Soc., vol. 126, no. 1, pp. 273–279, 2004.

[4] M. Ravichandran et al., "Cobalt ferrite nanowhiskers as T2 MRI contrast agent," RSC Adv, vol. 5, pp. 17223–17227, Feb. 2015.

[5] Tejabhiram Yadavalli, Hardik Jain, Gopalakrishnan Chandrasekharan, and Ramasamy Chennakesavulu, "Magnetic hyperthermia heating of cobalt ferrite nanoparticles prepared by low temperature ferrous sulfate based method," AIP Adv., vol. 6, no. 5, p. 55904, May 2016.

[6] S. Amiri and H. Shokrollahi, "The role of cobalt ferrite magnetic nanoparticles in medical science," Mater. Sci. Eng. C, vol. 33, no. 1, pp. 1–8, Jan. 2013.

[7] S. G. Kakade, R. C. Kambale, and Y.D. Kolekar, "Influence of pH on the structural and magnetic behavior of cobalt ferrite synthesized by sol-gel auto-combustion," AIP Conf. Proc., vol. 1665, no. 1, 2015.

[8] Niya Mary Jacob, Praveena Kuruva, Giridhar Madras, and Tiju Thomas, "Purifying water containing both anionic and cationic species using a (Zn, Cu) O, ZnO, and Cobalt Ferrite based multiphase adsorbent system," Ind. Eng. Chem. Res., vol. 52, no. 46, pp. 16384–16395, 2013.

[9] M. Kooti and M. Afshari, "Magnetic cobalt ferrite nanoparticles as an efficient catalyst for oxidation of alkenes," Sci. Iran., vol. 19, no. 6, pp. 1991–1995, Dec. 2012.

[10] Alex Goldman, Modern Ferrite Technology, Second. pittsburgh, PA, USA: Springer, 2006.

[11] R.S. de Biasi, C. Larica, A.B.S. Figueiredo, and A.A.R. Fernandes, "Synthesis of cobalt ferrite nanoparticles using combustion waves," Solid State Commun., vol. 144, no. 1–2, pp. 15–17, Oct. 2007.

[12] Praveena Kuruva, Shidaling Matteppanavar, S. Srinath, and Tiju Thomas, "Size Control and Magnetic Property Trends in Cobalt Ferrite Nanopaerticless Synthesized Using an Aqueous Chemical Route," IEEE, vol. 50, no. 1, Jan. 2014.

[13] Jae-Gwang Lee, Jae Yun Park, and Chul Sung Kim, "Growth of ultra-fine cobalt ferrite particles by a sol–gel method and their magnetic properties," J. Mater. Sci., vol. 33, no. 15, pp. 3965–3968, Aug. 1998.

[14] D. Sophia, M. Ragam, and S. Arumugam, "sysnthesis and characterisation of cobalt ferrite nanoparticles," Int. J. Sci. Res. Mod. Educ., 2016.

[15] Polina Yaseneva, Michael Bowker, and Graham Hutchings, "Structural and magnetic properties of Zn-substituted cobalt ferrites prepared by co-precipitation method," R. Soc. Chem., vol. 13, pp. 18609–18614, 2011.

[16] Tashihiko Sato, Tetsuo Iijima, Masahiro Seki, and Nobou Inagaki, "Magnetic properties of ultrafine ferrite particles," J. Magn. Magn. Mater., no. 65, pp. 252–256, 1987.

[17] Sichu Li, Limin Liu, V.T. John, C.J. O'Connor, and V.G. Harris, "Cobalt-ferrite nanoparticles: correlations between synthesis procedures, structural characteristics and magnetic properties," IEEE Trans. Magn., vol. 37, no. 4, Jul. 2001.

[18] N. Bogdanchikova, Alexey Pestryakov, M. H. Farias, Jesus Antonio Diaz, M. Avalos, and J. Navarrete, "Formation of TEM and XRD-undetectable gold clusters accompanying

big gold particles on TiO$_2$–SiO$_2$ supports," Solid State Sci., vol. 10, no. 7, pp. 908–914, Jul. 2008.

[19]  Wei-Chih Hsu, S.C Chen, P.C Kuo, C.T Lie, and W.S Tsai, "Preparation of NiCuZn ferrite nanoparticles from chemical co-precipitation method and the magnetic properties after sintering," Mater. Sci. Eng. B, vol. 111, no. 2–3, pp. 142–149, Aug. 2004.

[20]  G.A. El-Shobaky, A.M. Turky, b, N.Y. Mostafa, and S.K. Mohamed, "Effect of preparation conditions on physicochemical, surface and catalytic properties of cobalt ferrite prepared by coprecipitation," J. Alloys Compd., no. 493, pp. 415–422, 2010.

[21]  G.S. Shahane, Ashok Kumar, Manju Arora, R.P. Pant, and Krishan Lal, "Synthesis and characterization of Ni–Zn ferrite nanoparticles," J. Magn. Magn. Mater., no. 322, pp. 1015–1019, 2010.

[22]  Kanagaraj M, Sathishkumar P, Selvan G Kalai, Kokila I Phebe, and Arumugam S, "Structural and magnetic properties of CuFe$_2$O$_4$ as-prepared and thermally treated spinel nanoferrites," Indian J. Pure Appl. Phys., vol. 52, no. 2, pp. 124–130, Feb. 2014.

[23]  BP Ladgaonkar, CB Kolekar, and AS Vaingankar, "Infrared absorption spectroscopic study of Nd$^{3+}$ substituted Zn-Mg ferrites," Bull. Mater. Sci., vol. 25, no. 4, pp. 351–354, Aug. 2002.

[24]  Marykutty Thomas and K C George, "Infrared and magnetic study of nanophase zinc ferrite," Indian J. Pure Appl. Phys., vol. 47, Feb. 2009.

[25]  Raghvendra A. Bohara, Nanasaheb D. Thorat, Hemraj M. Yadav, and Shivaji H. Pawar, "One-step synthesis of uniform and biocompatible amine functionalized cobalt ferrite nanoparticles: a potential carrier for biomedical applications," New J. Chem., no. 38, pp. 2979–2986, Apr. 2014.

[26]  C. Nethravathi, Sonia Sen, N. Ravishankar, Michael Rajamathi, Clemens Pietzonka, and Bernd Harbrecht, "Ferrimagnetic Nanogranular Co3O4 through Solvothermal

Decomposition of Colloidally Dispersed Monolayers of α-Cobalt Hydroxide," J. Phisical Chem. B, vol. 109, no. 23, pp. 11498–11472, May 2005.

[27]   Z. P. Xu and H. C. Zeng, "Thermal evolution of cobalt hydroxides: a comparative study of their various structural phases," J. Mater. Chem., vol. 8, no. 11, pp. 2499–2506, 1998.

[28]   Jing Yang, Hongwei Liu, Wayde N. Martens, and Ray L. Frost, "Synthesis and Characterization of Cobalt Hydroxide, Cobalt Oxyhydroxide, and Cobalt Oxide Nanodiscs," J Phys Chem C, no. 114, pp. 111–119, 2010.

[29]   Shanmugam Yuvaraj, Lin Fan-Yuan, Chang Tsong-Huei, and Yeh Chuin-Tih, "Thermal Decomposition of Metal Nitrates in Air and Hydrogen Environments," J. Phys. Chem. B, vol. 107, no. 4, pp. 1044–1047, Jan. 2003.

[30]   M.A. Gabal, Reda M. El-Shishtawy, and Y.M. Al Angari, "Structural and magnetic properties of nano-crystalline Ni–Zn ferrites synthesized using egg-white precursor," J. Magn. Magn. Mater., vol. 324, no. 14, pp. 2258–2264, Jul. 2012.

[31]   E. Rezlescu, N. Rezlescu, C. Pasnicu, M.L. Craus, and D.P. Popa, "The influence of additives on the properties of Ni-Zn ferrite used in magnetic heads," J. Magn. Magn. Mater., vol. 117, pp. 448–454, 1992.

[32]   Nobuyoshi Koga, Shigeru Takemoto, Syougo Okada, and Haruhiko Tanaka, "A kinetic study of the thermal decomposition of iron(III) hydroxide oxides. Part 1. α-FeO(OH) in banded iron formations," Thermochim. Acta, vol. 254, pp. 193–207, Apr. 1995.

[33]   V. Pillai and D.O. Shah, "Synthesis of high-coercivity cobalt ferrite particles using water-in-oil microemulsions," J. Magn. Magn. Mater., no. 163, pp. 243–248, 1996.

[34]   Srikanth VV and Sreenivasulu KV, "Fascinating magnetic energy storage nanomaterials: A brief review," Recent Pat. Nanotechnol., vol. 10, 2016.

[35]   Donald R. Askeland and Pradeep P. Phulé, The Science and Engineering of Materials, 4th ed..

[36]     Lijun Zhao et al., "Studies on the magnetism of cobalt ferrite nanocrystals synthesized by hydrothermal method," J. Solid State Chem., vol. 181, no. 2, pp. 245–252, Feb. 2008.

[37]     Alberto Lopez-Ortega, Elisabetta Lottini, Cesar de Julian Fernandez, and Claudio Sangregorio, "Exploring the Magnetic Properties of Cobalt-Ferrite Nanoparticles for the Development of a Rare-Earth-Free Permanent Magnet," Chem. Mater., vol. 27, pp. 4048–4056, May 2015.

[38]     A.S. Ponce, E.F. Chagas, R.J. Prado, C.H.M. Fernandes, A.J. Terezo, and E. Baggio-Saitovitch, "High coercivity induced by mechanical milling in cobalt ferrite powders," J. Magn. Magn. Mater., vol. 344, pp. 182–187, Oct. 2013.

[39]     Francisco de Assis Olímpio Cabral, Fernando Luis de Araujo Machado, José Humberto de Araujo, João Maria Soares, Alexandre Ricalde Rodrigues, and Armando Araujo, "Preparation and magnetic study of the $CoFe_2O_4$-$CoFe_2$ nanocomposite powders," IEEE Trans. Magn., vol. 44, no. 11, Nov. 2008.

[40]     A. C. Lima, M. A. Morales, J. H. Araújo, J. M. Soares, D. M. A. Melo, and A. S. Carriço, "Evaluation of $(BH)_{max}$ and magnetic anisotropy of cobalt ferrite nanoparticles synthesized in gelatin," Ceram. Int., vol. 41, no. 9, pp. 11804–11809, Nov. 2015.G. Eason, B. Noble, and I. N. Sneddon, "On certain integrals of Lipschitz-Hankel type involving products of Bessel functions*," Phil. Trans. Roy. Soc.* London, vol. A247, pp. 529-551, Apr. 1955.